\documentclass{eas}
\usepackage{graphicx}
\def\solm{M$_{\odot}\,$}
\begin{document}
\title{Gas Dynamics in Central Parts of Galaxies} 
\author{Witold Maciejewski}\address{
INAF -- Osservatorio Astrofisico di Arcetri, Largo E. Fermi 5, 
50125 Firenze, Italy, \\ and Obserwatorium Astronomiczne Uniwersytetu
Jagiello{\'n}skiego, Cracow, Poland}
%\today
\begin{abstract}
Asymmetries in galactic potentials, either self-induced, or caused by 
a passing companion, play an important role in global gas dynamics in
galaxies. In particular, they are able to trigger gas inflow, which in 
turn feeds nuclear activity. In the inner kiloparsec, the inflowing gas 
becomes subject to various resonances induced by asymmetry in the 
potential. This produces complicated gas morphology and dynamics seen in 
the nuclear rings and spirals. Formation of nuclear rings is related
to barred galaxies, and explained by the orbital structure in bars.
One can get basic understanding of weak nuclear spirals from the linear 
wave theory, but stronger spirals are likely to be
out of the linear regime, and they appear as spiral shocks in 
hydrodynamical models. In addition, nuclear stellar bars are observed 
in a considerable fraction of disc galaxies, often nested inside the 
large bar; their orbital structure 
further modifies the nuclear gas dynamics. If a massive black hole (MBH) 
is present in the galactic centre, it governs the resonances even beyond 
its classically defined sphere of influence. Since resonances shape gas
dynamics in the nuclear region, the central black hole should be able 
to regulate gas flow around itself.
\end{abstract}
\maketitle
\section{Introduction}
Nuclear starbursts and Active Galactic Nuclei (AGN) have to be fed by mass 
transport into the active regions. Due to dissipation, gas inflow is easiest 
to trigger. There are two obvious dynamical mechanisms causing gas inflow in 
galaxies: interactions and asymmetries in galactic potentials. The first one
involves violent and transient phenomena with strong gas inflow towards the 
centres of the interacting galaxies seen in numerical simulations (see e.g.
Mihos \& Hernquist 1996). Closer analysis of the modeled merging sequence
shows that the inflow process has two stages: first, tidal perturbations
destabilize stellar discs, and then, each disturbed stellar disc exerts 
torques on the gas, and triggers inflow. Thus even in systems apparently as 
order-less as merging galaxies, gas dynamics is governed by the perturbed 
potentials of each individual galaxy. 

Here I review the dependence of gas dynamics on the gravitational potential 
in isolated galaxies, where it can be seen more clearly, since these galaxies
remain roughly unchanged on timescales corresponding to the formation of 
steady gas flow patterns. Inflow in isolated galaxies can occur steadily, 
which may in principle provide continuous feeding of the central
activity. However, the presence and the positions of dynamical resonances 
determine the mode of gas flow in the nuclear region: it may be quite complex, 
as the recent high-resolution observations of galactic centres indicate. 
Structured dust lanes, nested bars, as well as nuclear rings and spirals have 
been observed (Regan \& Mulchaey 1999, Martini \& Pogge 1999, Pogge \& Martini 
2002, Erwin \& Sparke 2002).

First, I briefly review global gas flow patterns in the most common asymmetry 
in disc galaxies, i.e. in the presence of a stellar bar (\S2), and then,
in subsections of \S3, I explore various modes of gas flow in the nuclear 
region. I also attempt to determine which modes may trigger significant 
inflow capable of feeding the AGN. Placing a Massive Black Hole (MBH) at 
the galaxy centre may add, remove or shift resonances, therefore such a 
black hole may regulate gas flow around itself. Finally, I make a few 
remarks on numerical methods (\S4).

\section{Gas dynamics in a barred galaxy outside the nuclear ring}
Galactic discs commonly develop stellar bars, and the majority of
observed galaxies is barred (e.g. Eskridge et al. 2000). The most 
established observational notion about gas morphology in barred galaxies 
is the existence of two straight or curved dust lanes in the main body of 
the stellar bar, which are symmetric around the galaxy centre, and slightly 
tilted to the bar's major axis (see e.g. NGC 1097, NGC 1300, NGC 4303, and
NGC 6951 in the Hubble Atlas of Galaxies, Sandage 1961). Strong 
discontinuities in gas velocity, indicative of shocks, have been measured 
along the dust lanes (Regan et al. 1997). This morphology and kinematics have
been seen in numerical models of gas flow in a fixed potential of the bar 
since the work of S{\o}rensen et al. (1976). It has been explained in terms 
of stable periodic orbits by Athanassoula (1992). 

There are two major families of stable prograde periodic orbits inside the 
bar: the $x_1$ orbits parallel to the bar, and the $x_2$
orbits orthogonal to it (Contopoulos \& Papayannopoulos 1980). The $x_2$ 
orbits are present only in galaxies with sufficient central mass 
concentrations and moderate bar pattern speeds --- in linear analysis this 
condition corresponds to the existence of an Inner Lindblad Resonance (ILR, 
see Binney \& Tremaine 1987, and \S3). By definition, the $x_2$ orbits exist 
only inside this resonance. Representative orbits of each family are plotted 
in Figure 1.

\begin{figure}
\vspace{-5cm}
\includegraphics[width=13cm]{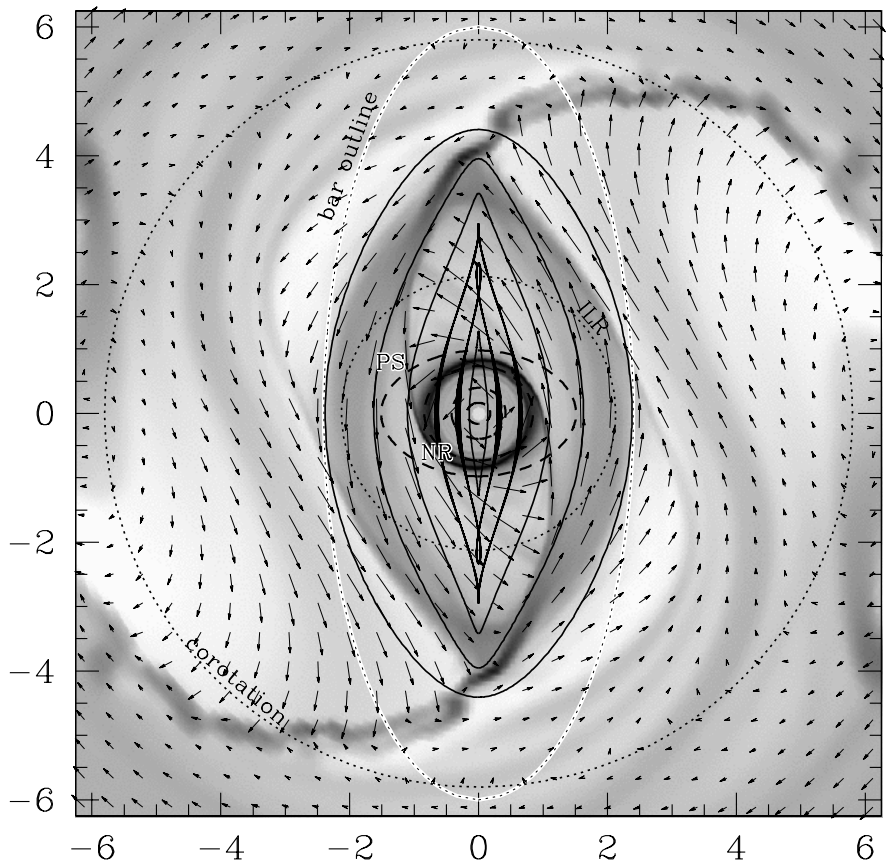}
\vspace{-2.8cm}
\caption{A representative snapshot of the gas density and velocity field in a 
barred galaxy, taken after 6 rotation periods of the bar, once the main flow 
patterns have been established. The gas, treated as a non-selfgravitating, 
isothermal fluid with a sound speed of 5 km/s, responds to a fixed 
gravitational potential 
of a bar, disc and spheroid, and is modeled with an Eulerian code on a fixed 
grid. The density is shown in grayscale, and arrows mark gas velocity in the 
reference frame rotating with the bar. Dotted circles mark corotation 
and the ILR. Examples of $x_1$ and $x_2$ orbits are drawn with solid 
and dashed lines, respectively. {\bf PS} marks the principal shock in 
the bar, {\bf NR} is the nuclear ring. Units on axes are in kpc. From the 
model S05 by Maciejewski et al. (2002)}
\end{figure}

In the outer parts of the bar, gas tends to follow the $x_1$ orbits. Further 
inwards, the $x_1$ orbits develop cusps and loops: stars can follow such 
orbits, but gas on self-intersecting orbits gets shocked. A robust feature in 
the majority of numerical models is a pair of principal shocks that develop 
on the leading side of the bar (marked PS in Fig.1). There, torques transfer 
angular momentum from the gas to the bar, and the gas falls towards the centre. 
Numerical models indicate considerable gas inflow in the principal shocks: 
up to 1 \solm yr$^{-1}$ for typical gas densities in the disc (Athanassoula 
1992; Regan et al. 1997). The shocked gas usually does not flow straight
to the galaxy centre: it tends to settle on the $x_2$ orbits, which have lower
value of the Jacobi integral $E_J$ (corresponding to the energy in the 
rotating frame). Consequently, the straight shocks are slightly tilted to the 
bar's major axis. If there were no $x_2$ orbits, straight shocks would develop
on the bar's major axis, possibly extending directly to the centre of the galaxy 
(Athanassoula 1992, Piner et al. 1995). Only a couple of galaxies with such 
central (on-axis) straight shocks have been observed (Athanassoula 1992), 
which suggests that most barred galaxies possess the ILR. The gas settling on
$x_2$ orbits, well inside the ILR (see Fig.1), often accumulates in
nuclear rings, where high density and no shear make favorable conditions
for star formation (Piner et al. 1995). Star-forming nuclear rings have
been observed in many galaxies (e.g. NGC 1512, Maoz et al. 2001; NGC 4314, 
Benedict et al. 2002). Nuclear rings may also form in a potential lacking
an ILR, given that this potential is almost axisymmetric at the nucleus
(Wada, priv. comm.).

\section{Gas dynamics around resonances inside the nuclear ring}
At resonances, the orbital frequency (Doppler-shifted to the reference frame 
of the bar) is commensurate with the frequency of radial oscillations. Thus 
a particle (or a gas cloud) at the resonance is subject to a monotonic force, 
which moves it away from there. The influence of the ILR is the strongest:
it creates the principal shocks described in \S2. For a bar rotating with 
a pattern speed $\Omega_B$, the ILR occurs where the orbital frequency 
$\Omega$ and the frequency of radial oscillations $\kappa$ are related by 
$\Omega - \Omega_B = \kappa/2$. Since $\Omega_B=$const, it is convenient to
rewrite this condition as $\Omega - \kappa/2 = \Omega_B$, where 
$\Omega - \kappa/2$ is called the frequency curve.

\subsection{The Triple ILR scenario}
The frequency curve always declines towards zero at large radii. It
goes to $+\infty$ at the galactic centre if a central MBH is present,
but even without a central MBH it can reach high values (larger than 
$\Omega_B$) at the nucleus, if the galaxy mass is sufficiently centrally 
concentrated. Thus, in such galaxies one should expect at least one ILR. 
However, if a galaxy has a central MBH, but a low central mass concentration
otherwise, the resonance curve may not be monotonic, and 3 ILRs can occur 
(Fig.2). They are called (from the innermost out) the nuclear ILR, the 
inner ILR and the outer ILR.

\begin{figure}
\vspace{-5mm}
\includegraphics[width=12cm]{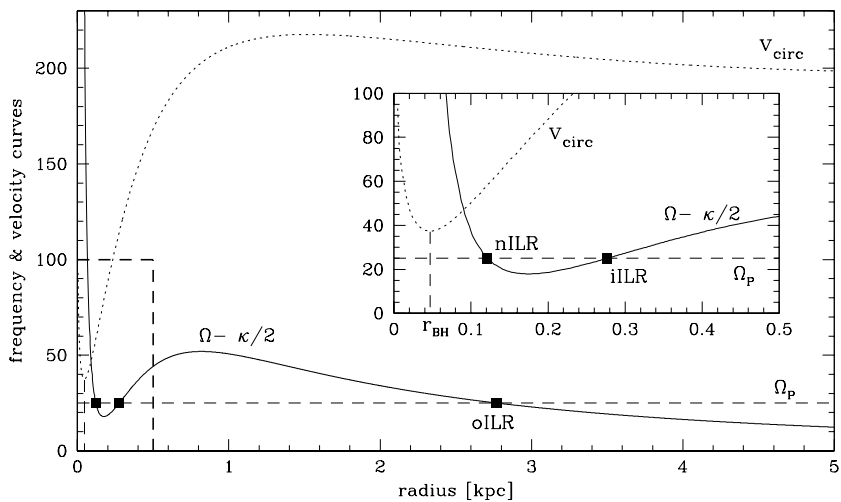}
\vspace{-5cm}
\caption{Velocity and frequency curves for a potential with 3 ILRs.
The circular velocity ($v_{\rm circ}$) is drawn with dotted line, the
frequency curve is solid. The constant pattern speed $\Omega_P$ is marked
with the horizontal dashed line. Positions of the three ILRs: nuclear (nILR),
inner (iILR), and outer (oILR) are marked with squares. A close-up of the
inner region is shown in the insert, where the sphere of influence of the
$10^7$\solm MBH has roughly the radius $r_{\rm BH}$.}
\end{figure}

The nuclear ILR is essentially related to the presence of the central MBH.
Fukuda et al. (1998, 2000) showed that gas flow around this resonance 
closely resembles the flow around the outer ILR described in \S2.
This gas morphology may prove essential in finding central MBHs in galaxies:
as can be seen in Figure 2, the nuclear ILR is located at the radius a few
times larger than the MBH's radius of influence (minimum of the rotation 
curve). Therefore we can learn about the central MBH from the gas morphology 
at radii larger than the ones useful in the conventional methods of stellar 
kinematics. On the other hand, shocks around the nuclear ILR strongly disturb 
gas kinematics; in this case, the assumption of a circularly rotating disc,
common when MBH mass is derived from kinematics of ionized gas (e.g. Macchetto
et al. 1997), may no longer be valid.

Because of the small size of the ring inside the nuclear ILR, the gas 
accumulating there may reach very high densities, so that self-gravity 
in gas will be the dominating force.
Fukuda et al. (2000) investigated the evolution of such a self-gravitating
ring, and showed that strong inflow induced by clumps in self-gravitating
gas will occur. This is one of the viable mechanisms of feeding the AGN;
here the central MBH enhances inflow onto itself due to the resonance that
it generates. Another possible scenario involves efficient star formation
in the ring, which leads to a nuclear starburst driven by the MBH.

Gas dynamics at the inner ILR is considerably different from that at the 
nuclear ILR and at the outer ILR. The orbital transition is 
different there as well: at the outer ILR, gas moves from the $x_1$ orbits 
with higher Jacobi integral $E_J$ to the nearly-circular $x_2$ ones with 
lower $E_J$. Inside the inner ILR, gas has to populate the $x_1$ orbits, 
which may be highly elongated at the centre (see Fig.1). The transition from
the flow along the $x_2$ orbits to the one following the $x_1$ family creates
a {\it leading} spiral at the inner ILR (Wada 1994, Knapen et al. 1995). 
The dynamics of this spiral is markedly different 
from the trailing spirals that form at the nuclear and outer ILRs:
it reflects a converging flow rather than a shock, with the hot gas upstream 
from the arm (Wada \& Koda 2001).

\begin{figure}
\vspace{-5mm}
\includegraphics[width=13cm]{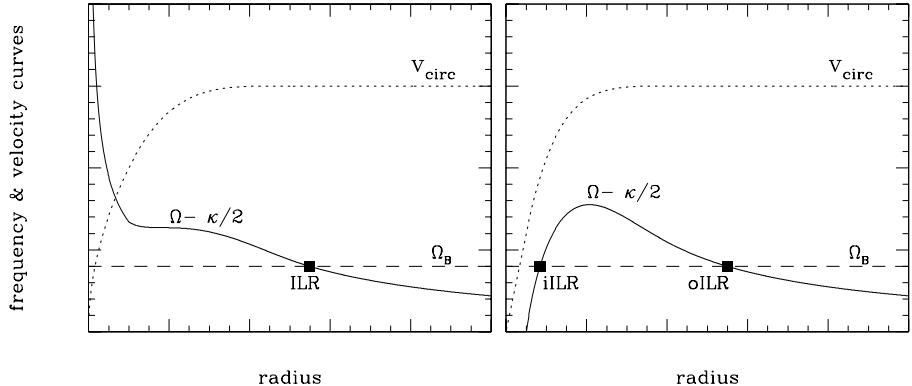}
\vspace{-78mm}
\caption{Two sample rotation curves (dotted): both are flat at large radii,
but the one on the left shows power-law rise at the centre, while the one on 
the right rises linearly. Solid line marks the corresponding frequency curves 
with positions of resonances (marked as in Fig.2) for a given pattern speed 
$\Omega_B$.}
\end{figure}

\subsection{Central rotation curves and mass profiles in galactic centres}
The triple ILR scenario predicts reversal of the spiral structure in the 
central parts of galaxies. Going inwards from the trailing spiral arms and the
principal shocks in the bar, one should encounter the leading spiral at
the inner ILR, and then a trailing one again in the innermost regions. This 
is not a commonly observed structure (\S3.3), which undermines the viability 
of the frequency curve in Figure 2. Furthermore, fits to surface brightness 
profiles of disc galaxy nuclei
show a very good agreement with power-law profile (Balcells et al. 2001).
For a power-law luminosity profile, the rotation curve is also a power
law (with positive exponent), as is the frequency curve (negative exponent 
here), which means that the latter rises monotonously inwards, and therefore 
has one ILR only.

The frequency curve is a derivative of the rotation curve, and is therefore
more sensitive to the variation of the rotation curve's shape than to its
value.  Consider two rotation curves (Fig.3), both flat at large radii, but
one a power-law in the inner part, while the other one has a linear inner
rise.  Although they both may fit a given set of observed data equally
well, they generate diametrically different frequency curves.  The rotation
curve with an inner power-law results in an inner power-law frequency
curve, which rises monotonically inwards (innermost radii in left panel of
Fig.3).  This implies one ILR only.  On the other hand, a linearly rising
rotation curve indicates solid body rotation, for which the frequency curve
is identically equal to zero (innermost radii in right panel of Fig.3).
This forces an additional ILR on the system, and now both outer ILR and
inner ILR are present.  Therefore, a linear rise of the central rotation
curve, although a good approximation to the data, may not properly reflect
the mass distribution in the inner parts of galaxies, or the resonances
that it induces.  Special care has to be taken when the presence and the
position of resonances are deduced from the observed rotation curve.
Resolution effects, like beam smearing, give the impression of a linear
rise in the central parts, which unavoidably implies the presence of an
inner ILR.

\subsection{Gas density waves within a single ILR}
Recent high-resolution maps of galactic centres reveal intricate dust
structures, which are often organized in a spiral pattern (Regan \& Mulchaey 
1999, Pogge \& Martini 2002). In disc galaxies,
these {\it nuclear spirals}, often winding by more than a $2\pi$ angle,
seem to connect smoothly to the outer spiral arms or to the principal dust 
lanes in the bar. This makes an indirect argument against the presence of 
an inner ILR, since theory and modeling indicate the reversal of the spiral 
there (\S3.1). On the other hand, the standard model of gas flow in the bar 
(\S2) does not account for nuclear spirals either: gas settles there in the
nuclear ring. Models by Athanassoula (1992) show curling of the inner parts
of the principal shock, but the resolution is too low there to follow this 
feature further inwards. 

\begin{figure}[b]
\vspace{-5mm}
\includegraphics[width=125mm]{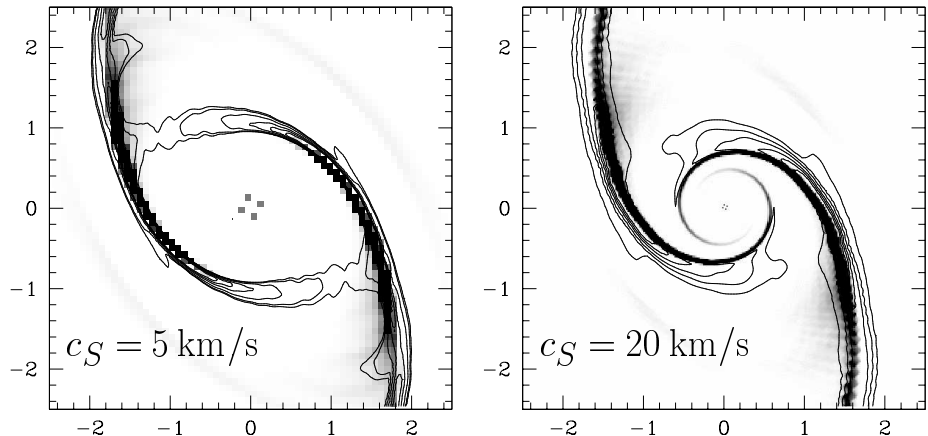}
\vspace{-12cm}
\caption{{\it Left:} A snapshot of gas flow with density marked by contours, 
and $div^2{\bf v}$ for $div{\bf v}<0$ (shock indicator) in grayscale for 
model S05 of cold gas 
from Maciejewski et al. (2002) at 0.1 Gyr, when gas leaving one principal 
shock reaches the opposite one. {\it Right:} A snapshot at the same time for 
model S20 of hot gas. Contours are at 1.7 (1.4 right), 2, 3, 5, 7 and 9 times
the initial gas density. Units on axes are in kpc.}
\end{figure}

For a given potential, gas flow depends 
sensitively on one parameter only: the velocity dispersion in the gas, 
which for an isothermal gas is represented by its temperature (Englmaier \& 
Gerhard 1997; Patsis \& Athanassoula 2000). Gas dynamics in the nuclear 
region changes remarkably with the sound speed in the isothermal gas ($c_S$): 
cold gas ($c_S \sim 5$ km/s) settles on the nuclear ring, as described in 
\S2, while hot gas ($c_S > 10$ km/s) forms a 
nuclear spiral (Maciejewski 1998). In the 
evolutionary sequence of gas flow in the bar, the principal shocks form first 
as trailing spirals, and then straighten as the bar grows in strength, with 
their inner parts curling inwards (Fig.4). In hot gas, the curling shock 
from one side of the bar never meets its counterpart from the opposite side, 
and both of them propagate towards the centre as an $m$=2 spiral mode. 
The convergence of the flow in such created nuclear spiral is high
(large negative $div{\bf v}$), which
is indicative of a shock. Significant gas inflow to the galactic centre may 
occur if the spiral shock proves to be efficient in transferring angular 
momentum from the gas to the bar. On the other hand, the principal shock in 
cold gas ceases to propagate inwards. Gas emerging from its innermost 
parts remains on a relatively open trajectory (Fig.4), and hits the shock 
on the opposite side of the bar. Hydrodynamical interaction of the flow 
with the shock results here in gas settling into the nuclear ring.

A good grasp of the dynamics of nuclear spirals can be gained when one
considers them as density waves generated by gravitational torques in a 
non-selfgravitating gas (Englmaier \& Shlosman 2000). In the linear 
approximation, the dispersion 
relation for the $m$=2 mode allows for wave propagation inside the
corotation only when $\Omega - \kappa/2 > \Omega_B$, i.e. between
the outer ILR and the inner ILR (if present) or the galactic centre.
Since nuclear spirals are observed in disc galaxies down to the HST
resolution limit, the inner ILR must reside inside that radius in these 
galaxies, if it is present at all.

Although some nuclear spirals in strongly barred galaxies may be well out of 
the linear regime,
the density-wave mechanism applies to a weak spiral inside the nuclear ring
seen in the models by Maciejewski et al. (2002). It is also extremely 
efficient in very weak bars, where gas dynamics closely follows the linear 
theory. Its predictions are the same for any weak nonaxisymmetry of the 
potential, including lopsidedness, because low harmonics of the mass 
distribution play a role there. Maciejewski (2001) studied gas flow in the 
gravitational potential with the quadruple moment of the bar 10 times lower, 
and the bar's axis ratio half that of the standard 
model presented in \S2. Such a bar is likely to be
too weak to be detected observationally, and it does not generate principal
shocks of any shape. Nevertheless, a clear spiral density enhancement
is seen inside the ILR. In accordance with the density-wave theory, it
extends all the way to the galactic centre if there is no inner ILR, or
ceases at the inner ILR if it is present. Unlike nuclear spirals in 
strong bars, which unwind outwards rapidly in order to match the principal
shock in the bar, nuclear spirals in very weak bars follow the linear
mode longer, and wind up to 3 times around the centre (6$\pi$ angle).
Thus tightly wound nuclear spirals may be observationally associated
with galaxies where no bar has been detected.

\subsection{Sorting out distinct scenarios of gas flow in nested bars}
Galaxies possessing nested bars (a secondary bar inside the main one)
have been noticed already almost 30 years ago (de Vaucouleurs 1974), but 
only recent studies (Laine et al. 2002, Erwin \& Sparke 2002) showed them 
to be quite a common phenomenon: up to 30\% of barred galaxies contain 
nested stellar bars. The random relative orientation of the bars suggests
that they rotate independently. For a while nested bars became a favourite 
mechanism to fuel the nucleus: if gas flow in the secondary bar is analogous 
to the one in the main bar, then the secondary bar may force gas deeper 
into the potential well. Meanwhile, the pioneering work of Shlosman et al. 
(1989) examined fueling mechanism that involves instability in the 
self-gravitating nuclear gaseous disc. As the gas inflow in the large bar 
stagnates at the nuclear ring or disc, gas accumulates there, and can 
become gravitationally unstable. By analogy to instability of the 
galactic stellar disc, Shlosman et al. proposed that a {\it gaseous} 
inner bar forms inside the large bar (gaseous bar scenario).

The gaseous bar scenario may be related to the
observed nested {\it stellar} bars, when star
formation occurs in the dense gaseous inner bar, or when the inner
gaseous bar drags stars of the nuclear component (Combes 1994). Neither
of these links is likely though, since the gaseous bar scenario
relies on {\it global} instability in self-gravitating gas,
while cold massive nuclear discs are more likely to become unstable
{\it locally}, and form clumps. This is clearly seen in simulations
by Heller \& Shlosman (1994): although the nuclear gaseous disc first
collapses into a bar-like shape, clumps develop very quickly, so that
no bar is seen a few Myr after its formation. This time is too short 
to form stars or to alter stellar dynamics, and the star-forming clumps 
do not follow bar-like orbits either. Therefore nested {\it 
stellar} bars are unlikely to be produced by the gaseous bar scenario.
However, this scenario may lead to an episodic gas inflow: one galaxy 
where it may operate is Circinus (Maiolino et al. 2000) --- one of the 
three closest AGN. A strong gas inflow
on a scale $\sim$50 pc is seen on one side of the nucleus, but it is
not accompanied by any outflow. It is likely that the unstable gaseous disc
developed an $m=1$ instability mode there, and is feeding the nucleus
in a short episode.

In order to estimate the dynamical significance of the gaseous component
Friedli \& Martinet (1993) and Combes (1994) looked for formation of
nested bars in numerical simulations involving stellar and gaseous particles. 
They were able to create secondary {\it stellar} bars, which decouple from
the large bars after the central mass concentration increases substantially
as a result of gas inflow in the large bar. These inner stellar bars can then 
drag gas with them. In this mechanism, the role of gas is to increase central
mass concentration, and to prevent heating of the central stellar component.
No gravitational instability in the gaseous disc is involved, contrary to
the mechanism proposed by Shlosman et al. (1989). Note however that
these simulations use numerical methods suffering from 
excessive gas inflow to the centre (see \S4), and therefore central 
mass concentrations may be unrealistically high, which undermines their 
conclusion. Also, the size ratio of the bars produced in this mechanism
(Set III in Friedli \& Martinet 1993) is 2 --- much lower than the observed
values. It indicates that these models may miss the physical mechanisms
operating in the observed systems. A more realistic size ratio (3.8)
is produced in Set II of Friedli \& Martinet, but there the gas only
prevents stars from heating, and the initial conditions are particularly
chosen so that two bars are formed.

Since inner bars are often observed in near infrared, they are 
most likely made out of old stellar populations, and are long-lived.
They are often devoid of gas, hence it is likely that purely stellar-dynamical
mechanisms forming decoupled nested bars are in operation there
(Rautiainen \& Salo 1999). Maciejewski \& Sparke (2000) explored limitations
which are imposed on nested stellar bars by their orbital structure.
Since orbits in a potential of two independently rotating bars (a class
of pulsating potentials) do not close in general, they generalized the
concept of the orbit to pulsating potentials, and found sets of particles
populating closed curves that support bars in their motion. If resonant 
coupling is assumed to minimize the number of chaotic zones (Sygnet et al. 
1988), then curves hosting sets of particles that support
the inner bar are round and end well within its corotation, so that no 
shocks in gas are induced. Maciejewski et al. (2002) confirmed this with
hydrodynamical simulations: gas flow in the inner stellar bar, which is in 
resonant coupling with the large bar, does not develop straight shocks, and 
does not increase the inflow. It instead organizes itself in an ellipsoidal 
pattern around the inner bar, like the one seen in color maps of NGC 3081
(Regan \& Mulchaey 1999). Nevertheless, nested bars may not follow the 
resonant coupling assumed in this scenario, and other modes of
gas flow in the inner bar are possible, including straight dust lanes
(Maciejewski 2002).

\section{Remarks about numerical methods}
Numerical methods used in modeling of gas flow in galaxies can generally 
be divided into 3 classes: one treats gas as a set of ballistic clouds 
with some rules for energy and momentum exchange (sometimes called 
'sticky particles'), the second solves hydrodynamical equations 
involving the SPH algorithm (Smooth 
Particle Hydrodynamics), while the third solves these equations directly on 
the grid. The first method, and to some extent the second, involves a number 
of free parameters in gas description, which can be adjusted to describe 
particular features of the ISM. On the other hand, the third method may 
involve algorithms (like the Riemann solver) which limit the set of
free parameters to purely numerical ones, and which attempt to give the best 
approximation to the dynamics of non-viscous gas. This leads to unique
hydrodynamical solutions, but it remains unclear how the real ISM differs 
from non-viscous fluid.

Consequences of numerical viscosity are particularly significant in 
studies of gas inflow. Bulk viscosity in SPH-based codes is necessary to 
stabilize post-shock oscillations, and it is quite large.
One can compare the evolution of density within the inner 300 pc (inside 
the nuclear ring) in the grid-based model (Piner et al. 1995) to that in 
the SPH model (Patsis \& Athanassoula 2000) for the same potential: in the 
first model, the density remains roughly constant, while in the second one
it increases by more than a factor of 50 after 8 pattern rotations. Such a 
central mass concentration may be not realistic, and its strong dynamical 
influence, postulated by Friedli \& Martinet (1993) and others, may not
be relevant to present-day galaxies.
Also the importance of self-gravity in gas may need to be scaled down. Heller 
\& Shlosman (1994) measure mass of gas within 500 pc at some stage of their 
SPH simulations to be half of the dynamical mass. This is significantly
more than the observed values (see e.g. Sakamoto et al. 1999). They also
measure the mass of the central gas clump to be $3\times10^9$\solm, when
usually no more than $10^9$\solm of gas is seen in galaxy centres. In the SPH
simulations by Berentzen et al. (1998) a considerable amount of gas quickly 
falls onto the 'central accreting object', so that mass in gas accumulated 
there amounts to 1.6\% of the total galaxy mass. This is an order of
magnitude larger than the currently measured central MBH masses (Kormendy
\& Gebhardt 2001).

All 3 classes of methods discussed above usually represent the ISM as an 
isothermal fluid. Although this is a valid approximation (Cowie 1980), which 
has produced many significant results so far, one will eventually need a more 
realistic description of the ISM. Immense progress has been made here in the 
recent years by Wada \& Norman (1999, 2001), who developed a grid-based code
which allows for cooling and heating of the gas, as well as for self-gravity.
As a result, a multiphase medium forms, with lognormal density distribution
over 4 orders of magnitude. Shear by differential rotation seems to be
sufficient to create flocculent nuclear spirals, and dynamics of this
medium in presence of a weak bar has been studied recently (Wada \& Koda 2001).

\section{Conclusions}
Asymmetries in the galactic potential can alter gas dynamics, generate
inflow, and consequently speed up galaxy evolution. Gas dynamics in centres
of galaxies strongly depends on the presence of resonances created by
asymmetries in the potential. The presence and the number of resonances 
is regulated by the central mass distribution. In particular, the MBH
in the centre of galaxy can create and/or move the resonances, and 
thus regulate gas flow around itself. Grand design nuclear spirals 
observed in centres of galaxies are consistent with the presence of one
ILR only. Some of them may be sites of shocks that increase inflow to
the galactic centre.

Two fundamentally different gas flows have been labeled as 'gas flow in
nested bars', which often causes confusion. One mode of the flow,
originally proposed by Shlosman et al. (1989), refers to gaseous inner
bars, which form in unstable gaseous nuclear discs. In this scenario
the bar {\it is} the flow: it involves rapid radial gas motion on
timescales of the order of dynamical time. It may be able to feed the AGN, 
but modeling work was not able yet to find connection between it and the 
nested stellar bars that we observe in galaxies. The second scenario refers 
to the flow in nested
stellar bars. Stable orbits supporting such systems enable their
long lifetime, and at the same time they restrict gas flow to follow
elliptical trajectories (Maciejewski \& Sparke 2000). In this scenario 
nested bars do not increase inflow to the galactic centre.

\vspace{5mm}

{\bf Acknowledgments.} I would like to thank the organizers of the
Galactic Dynamics Workshop at the JENAM 2002 Symposium for their invitation
to give this review. Insightful comments by Lia Athanassoula, Eric 
Emsellem, Peter Erwin, Roberto Maiolino, Alessandro Marconi, Eva 
Schinnerer, Peter Teuben and Keiichi Wada have been invaluable to
this review.

%%-----------------------------
%%      your bibliography
%%-----------------------------

\end{document}